\documentclass[aps,prd,onecolumn,groupedaddress,showkeys,nofootinbib,amssymb]{revtex4}

\usepackage[T1]{fontenc}
\usepackage[latin1]{inputenc}
\usepackage{graphicx}
\usepackage[english]{babel}
\usepackage{amsmath}
\usepackage{amssymb}
\usepackage{amsfonts}

\begin{document}
\def\pp{{\, \mid \hskip -1.5mm =}}
\def\cL{{\cal L}}
\def\beq{\begin{equation}}
\def\eneq{\end{equation}}
\def\bea{\begin{eqnarray}}
\def\enea{\end{eqnarray}}
\def\tr{{\rm tr}\, }
\def\nn{\nonumber \\}
\def\e{{\rm e}}

\title{\textbf{  New Spherically Symmetric Solutions in $f(R)$-gravity by Noether Symmetries}}

\author{Salvatore Capozziello$^{1,2}$\footnote{e-mail address: capozziello@na.infn.it}, Noemi Frusciante$^{3,4}$\footnote{e-mail address: noemi.frusciante@sissa.it}, Daniele Vernieri$^{3,4}$\footnote{e-mail address: daniele.vernieri@sissa.it}}

\affiliation{\it $^1$Dipartimento di Scienze Fisiche,
Università di Napoli {}``Federico II'' and  $^2$INFN Sez. di Napoli,
Compl. Univ. di
Monte S. Angelo, Edificio G, Via Cinthia, I-80126, Napoli, Italy \\
$^3$SISSA-International School for Advanced Studies, Via Bonomea 265,
34136 Trieste, Italy and 
$^4$INFN, Sez. di 
Trieste, Via Valerio 2, 34127, Trieste, Italy}

\date{\today}

\begin{abstract}
Spherical symmetry for  $f(R)$-gravity is discussed by searching for Noether symmetries. The method consists in 
selecting  conserved quantities in  form of  currents that reduce  dynamics of  $f(R)$-models compatible with symmetries. In this way we get a general method to obtain constants of motion without setting {\it a priori}  the form of  $f(R)$.  In this sense,  the Noether symmetry results a  physical criterium. Relevant cases are discussed.

\end{abstract}

\keywords{alternative theories of gravity; exact solutions; Noether symmetries}
\maketitle

\section{Introduction}

The issue to renormalize General Relativity (GR)
leads to effective theories of gravity  including corrections in the curvature invariants to the
Einstein-Hilbert action
\cite{renorma1,renorma2,renorma3,renorma4}.  In addition, the  discovery
that the expansion of the universe is accelerated \cite{SN} has
led to the introduction of  the so-called {\it dark energy} to explain the accelerated expansion of the Hubble fluid
 which, up to now, has no agreed interpretation at fundamental level.
 
 In order to avoid this {\em ad hoc} ingredient, it has been proposed \cite{curvature} 
 that, perhaps, one needs to modify GR
at  the cosmic  scales to address the observational evidences.

In particular, $f(R)$-gravity, where $f$ is a  function of the Ricci scalar $R$ (in particular an analytic function),  has been considered to
explain the present cosmic acceleration without dark energy
(see \cite{Odi, CapFra, SotiriouFaraoni10, DeFeliceTsujikawa10}
for reviews). Here we focus on the metric version of these
theories, where the field equations  are
of fourth order for the metric.

The full investigation of such theory is the first step toward the so-called {\it Extended Theories of Gravity} where effective actions involving
generic curvature invariants are taken into account \cite{mauro1,mauro2,book,book1,report}. The goal is to address, by gravitational sector, 
problems like dark energy and dark matter at infrared scales and renormalization at ultraviolet scales.
 
 In this view, exact solutions of the field equations are important in order to
gain  insight into the mathematical and physical content of a
theory \cite{bk:dk}. They can be obtained only upon assuming
special  symmetries. In the case of metric $f(R)$-gravity  {\em in
vacuo}, assuming spherical symmetry does not automatically lead to
the Schwarzschild or Schwarzschild-de Sitter solutions
\cite{spherical}. The latter do solve the $f(R)$ vacuum field
equations, but the Birkhoff theorem of GR
does not hold in metric $f(R)$ gravity (see
\cite{FaraoniJebsen10,artu} for a detailed discussion) and,
therefore, these are not all the solutions within the family of spherically symmetric solutions \cite{Multamaki, Multamaki1,Multamaki2,CliftonBarrow}. In fact, the class of
spherically symmetric solutions of metric $f(R)$-gravity is still
unexplored, with only a handful of analytical solutions beyond
Schwarzschild-de Sitter being known. When one considers axially
symmetric solutions, the situation is even worse: apart from  the
Kerr metric \cite{Kerr}, only the solutions of Ref.~\cite{axial}
are presently available.

Because of the fact that these models have fourth order field equations, it is usually  difficult to find exact solutions in full generality without imposing a-priori a particular form of the $f(R)$ function. 
To  this aim, it is possible to consider spherically symmetric background and  demonstrate that it is possible to find exact solutions via the Noether Symmetry Approach \cite{artu,lambiase,defelice,greci}. In fact, choosing an appropriate $f(R)$-Lagrangian, it is possible to find out conserved Noether currents  which will be useful to solve  dynamics. This approach is very powerful due to the fact that it allows to find a closed system of equations, where we do not need to impose the particular form of the $f(R)$  which, on the other hand,   is selected by the Noether symmetry itself.
In Section \ref{II}, we introduce the $f(R)$-gravity using a suitable Lagrangian approach in the hypothesis of a  spherically symmetric metric. Dynamical equations are solved in the case of constant curvature recovering the standard case of the Schwarzschild-de Sitter solution. In Section \ref{III}, we describe, in general, the Noether Symmetry Approach and in Section \ref{IV}, we apply it to spherically symmetric 
$f(R)$-gravity. In Section \ref{V} we find solutions to the dynamics for both constant curvature and in the hypothesis of a static metric. Conclusions are given in Section \ref{VI}.

\section{Spherically symmetry in  $f(R)$ gravity}     \label{II}

Let us start with a generalization of the Hilbert-Einstein action

\beq\label{1.0}
\mathcal{S}=\int{d^4x\sqrt{-g}f(R)},
\eneq
where \textit{f(R)} is an analytic function of the Ricci scalar $R$ and $g$ is the determinant of the metric $g_{\mu \nu}$. We consider a time dependent spherically symmetric metric 

\beq\label{1.1}
ds^2=e^{\nu(r,t)}dt^2-e^{\lambda(r,t)}dr^2-r^2d\Omega^2,
\eneq
where $\nu(r,t)$ and $\lambda(r,t)$ are unknown functions of $r$ and $t$ and $d\Omega^2=(d\theta^2+sin^2\theta d\phi^2)$, using unit of $c=1$. In the following equations, we omit the $t$ and $r$ dependence, to avoid burdening the text. The Ricci scalar can be written as 

\beq\label{1.2}
R=-\frac{2}{r^2}+e^{-\lambda}\left[\nu_{rr}+\frac{1}{2}\nu_{r}(\nu_r-\lambda_r)+\frac{2}{r}(\nu_r-\lambda_r)+\frac{2}{r^2}\right]+e^{-\nu}\left[\frac{1}{2}\lambda_t(\nu_t-\lambda_t)-\lambda_{tt}\right],
\eneq
where  the subscripts $r$ and $t$  indicate   radial and time partial derivatives.
In order to obtain a canonical Lagrangian, function of $\nu(r,t)$, $\lambda(r,t)$ and their first derivatives, we use the method of Lagrange multipliers \cite{spherical,Zerbini,odilagrange}. Let us  introduce the   variable $\xi$ which is  the Lagrange multiplier: this allows   to set the Ricci scalar $R$ as a constraint for the dynamics. In this way the point-like action can be written as

\beq\label{1.3}
\tilde{\mathcal{S}}=\int drdt \,\,r^2e^{\frac{\nu+\lambda}{2}}\,\left\{f(R)-\xi\left[R+\frac{2}{r^2}-e^{\lambda}\biggl[\nu_{rr}
+\frac{1}{2}\nu_{r}(\nu_r-\lambda_r)+\frac{2}{r}(\nu_r-\lambda_r)+\frac{2}{r^2}\biggr]-e^{-\nu}\left[\frac{1}{2}\lambda_t(\nu_t-\lambda_t)-\lambda_{tt}\right]\right]\right\}.
\eneq
By this approach, we can consider $R(r,t)$, $\nu(r,t)$ and $\lambda(r,t)$ as independent Lagrangian variables. The Lagrange multiplier can be easily found by varying the action with respect to $R$, and the result is

\beq\label{1.4}
\xi=f'(R),
\eneq
where the prime denotes the derivative with respect to $R$. Using this expression for $\xi$ and integrating by parts, action (\ref{1.3}) can be rewritten as  

\beq\label{1.13}
\tilde{\mathcal{S}}=\int dtdr\,\, r^2e^{\frac{\nu+\lambda}{2}}\biggl\{f(R)-Rf'(R)-2\frac{f'(R)}{r^2}+e^{-\lambda}\left[2\frac{f'(R)}{r^2}-2\frac{f'(R)}{r}\lambda_r-f''(R)R_r\nu_r\right]+e^{-\nu}f''(R)R_t\lambda_t\biggr\}.
\eneq
Clearly, action (\ref{1.13}), when varied with respect to $R$, $\nu$, and  $\lambda$, gives the same equations of (\ref{1.0}). Furthermore, we are disregarding a divergence and this fact could affect the conserved current. Finally, we are going to consider internal (i.e. vertical) transformations so that it is not important to know the transformation rules of $\nu$ and $\lambda$ with respect to spacetime transformations.
The equations of motion are 

\bea
&-&\frac{2}{r}f''(R)R_r+\frac{1}{2}f''(R)\lambda_rR_r-f'''(R)R_r^2- f''(R)R_{rr} +\frac{1}{r}f'(R)\lambda_r-\frac{1}{r^2}f'(R)+\nonumber \\&+&e^{\lambda}\left[\frac{1}{2}Rf'(R)-\frac{1}{2}f(R) 
+\frac{1}{r^2}f'(R)\right]+\frac{1}{2}e^{-\nu}f''(R)R_t\lambda_t=0, \label{1.6}
\enea
for $\lambda$ and

\bea
&&e^{-\nu}\left[-\frac{1}{2}f''(R)\nu_tR_t+f'''(R)R_t^2+f''(R)R_{tt}\right] 
-\frac{1}{r^2}f'(R)-\frac{1}{r}f'(R)\nu_r-\frac{2}{r}f''(R)R_r-\frac{1}{2}f''(R)R_r\nu_r + \nonumber \\ &&+e^{\lambda}\left[\frac{1}{2}Rf'(R)-\frac{1}{2}f(R)+\frac{1}{r^2}f'(R)\right]=0, \label{1.7}
\enea
for $\nu$.
The equation for $R(r,t)$ is given by Eq. (\ref{1.2}), that is the Euler constraint on the dynamics.
Finally, we obtain a system of three partial differential Eqs. ((\ref{1.6}), (\ref{1.7}) and (\ref{1.2})) in the  unknown functions $\nu(r,t)$, $\lambda(r,t)$ and the constraint $R(r,t)$, once $f(R)$ is given.
As a first check it is easy to see that Eqs. (\ref{1.6}) and (\ref{1.7}) allow us to recover the well-known results of GR by replacing $f(R)=R$. \\
It is straightforward to show that when curvature is constant over space and time ($R=R_0$), an interesting solution can be found. Indeed subtracting Eqs. (\ref{1.6}) and (\ref{1.7}) we immediately obtain 

\beq\label{1.8}
\lambda_r+\nu_r=0,
\eneq
and substituting into Eq. (\ref{1.6}),  the solution for the metric component $g_{rr}$ is

\beq\label{1.9}
e^{\lambda(r,t)}=\frac{1}{1-\frac{\Lambda r^2}{3}-\frac{A_1(t)}{2r}},
\eneq
where $A_1(t)$ is an integration function depending on time, and 

\beq\label{1.10}
\Lambda=-\frac{1}{2}\left[R_0-\frac{f(R_0)}{f'(R_0)}\right].
\eneq
Solving Eq. (\ref{1.8}), we get

\beq\label{1.11}
e^{\nu(r,t)}=A_2(t)\left[1-\frac{\Lambda r^2}{3}-\frac{A_1(t)}{2r}\right]=A_2(t)e^{-\lambda(r,t)},
\eneq
where $A_2(t)$ is another integration function depending  on time. 
 Inserting Solutions (\ref{1.9}) and (\ref{1.11}) into Eq. (\ref{1.2}), we have a constraint on the integration functions: it is easy to show  that the only way to make Eq. (\ref{1.2}) consistent is to  choose  $A_1(t)=A_1=$constant, for which we recover the well-known static Schwarzschild-De Sitter solution

\beq\label{1.12}
ds^2=\left[1-\frac{\Lambda r^2}{3}-\frac{A_1}{2r}\right]d\tilde{t}^2-\frac{1}{1-\frac{\Lambda r^2}{3}-\frac{A_1}{2r}}dr^2-r^2d\Omega^2, \nonumber \\
\eneq 
where  time as been scaled accordingly 

\beq
d\tilde{t}^2=A_2(t)dt^2,    \label{1.14}
\eneq
being $R_0=-4\Lambda$.
The metric is  time independent, and we have thus proved the \textit{Birkhoff theorem} in the case of $f(R)$-gravity  with constant curvature. It is worth noticing that such a theorem does not  hold in general for any $f(R)$-model as  shown in \cite{book,artu}.

\section{Noether symmetry approach}         
 \label{III}

Let us consider now the Noether Symmetry Approach by which it is possible  to obtain conserved quantities asking for the invariance of the Lagrangian under the Lie derivative along an appropriate vector field. In this way, constraints on dynamics are achieved  and we it is  possible to  solve  the equations of motion. In general, a point-like canonical Lagrangian $\mathcal {L}$ depends on the variables $q^j(x^\mu)$ and on their derivatives $\partial_\nu q^j(x^\mu)$.  The corresponding Euler-Lagrange equations are

\beq
\partial_\mu\frac{\partial\mathcal {L}}{\partial\partial_\mu q^j}=\frac{\partial \mathcal{L}}{\partial q^j}.  \label{2.0}
\eneq
Contracting eq. (\ref{2.0}) with some unknown functions $\alpha^j=\alpha^j(q^i)$, it gives

\beq
\alpha^j\left(\partial_\mu\frac{\partial\mathcal {L}}{\partial\partial_\mu q^j}-\frac{\partial \mathcal{L}}{\partial q^j}\right)=0. \label{2.1}
\eneq
Since we can write

\beq
\alpha^j\partial_\mu\frac{\partial\mathcal {L}}{\partial\partial_\mu q^j}=\partial_\mu\left(\alpha^j\frac{\partial\mathcal {L}}{\partial\partial_\mu   \label{2.2} q^j}\right)-\left(\partial_\mu\alpha^j\right)\frac{\partial\mathcal {L}}{\partial\partial_\mu q^j},
\eneq
from Eq. (\ref{2.1}) we immediately obtain

\beq
\partial_\mu\left(\alpha^j\frac{\partial\mathcal {L}}{\partial\partial_\mu q^j}\right)=\alpha^j\frac{\partial \mathcal{L}}{\partial  \label{2.3} q^j}+\left(\partial_\mu\alpha^j\right)\frac{\partial\mathcal {L}}{\partial\partial_\mu q^j}=L_{\bf X}\mathcal{L},
\eneq
where $L_{\bf X}$ denotes the Lie derivative along the vector field

\beq
{\bf X}=\alpha^j\frac{\partial}{\partial q^j}+\left(\partial_\mu\alpha^j\right)\frac{\partial}{\partial\partial_\mu q^j}, \label{2.4}
\eneq
which represent the generator of symmetry.
We can immediately infer the Noether Theorem which states that
if $L_{\bf X}\mathcal{L}=0$,  the Lagrangian $\mathcal{L}$ is invariant along the vector field ${\bf X}$. As a consequence, we can define the current \cite{dan}

\beq
j^\mu=\alpha^j\frac{\partial\mathcal {L}}{\partial\partial_\mu q^j}\,,   \label{2.5}
\eneq 
which is conserved being

\beq
\partial_\mu j^\mu=0.   \label{2.6}
\eneq
As already developed in \cite{stabile}, the presence of Noether symmetries allows to reduce dynamics and then to find out exact solutions. Specifically, symmetries select also the analytic form of $f(R)$.
A similar approach has been developed for cosmological solutions \cite{defelice}.

\section{Noether symmetries in spherically symmetric $f(R)$ gravity}       
 \label{IV}

Let us take into account the  case of  spherical-symmetry in  $f(R)$-gravity. For the sake of generality, we can assume that the vector field ${\bf X}$ depends on the whole set of configuration variables that are  functions of the radial and time coordinates.
The configuration space is  $\mathcal{Q}=\{\nu(r,t), \lambda(r,t), R(r,t)\}$, so that the generator of symmetry\footnote{It is worth noticing that $R$, due to the constraint Eq. (\ref{1.2}), is not a proper variable. However, it is needed in order to make canonical the Lagrangian.} becomes:

\beq \label{2.7}
{\bf X}=\alpha\frac{\partial}{\partial \nu}+\beta\frac{\partial}{\partial \lambda}+\gamma\frac{\partial}{\partial R}+\alpha_r\frac{\partial}{\partial \nu_r}+
\beta_r\frac{\partial}{\partial \lambda_r}+\gamma_r\frac{\partial}{\partial R_r}+\alpha_t\frac{\partial}{\partial \nu_t}+\beta_t\frac{\partial}{\partial \lambda_t}+\gamma_t\frac{\partial}{\partial R_t},
\eneq
where $\alpha$, $\beta$ and $\gamma$ are functions of the variables $\nu$, $\lambda$ and $R$. 
As discussed above, a symmetry exists if the equation $L_{\bf X}\mathcal{L}=0$ has solutions for  the functions $\alpha$, $\beta$ and $\gamma$ where at least one of them is different from zero.
To verify such a condition, we can   set to zero the coefficients of  terms $\nu_r^2$, $R_r^2$, $\lambda_t^2$, $R_t^2$, $\nu_r\lambda_r$, $\nu_r R_r$, $\lambda_r R_r$, $\nu_t\lambda_t$, $\nu_t R_t$ and $\lambda_t R_t$ in $L_{\bf X}\mathcal{L}=0$ \cite{cimento}. We obtain  the following system of partial differential equations, linear in $\alpha$, $\beta$ and $\gamma$:

\beq
-r^2e^{\frac{\nu-\lambda}{2}}f''(R)\frac{\partial\gamma}{\partial \nu}=0,     \label{2.8}
\eneq

\beq
-r^2e^{\frac{\nu-\lambda}{2}}f''(R)\frac{\partial\alpha}{\partial R}=0,         \label{2.9}
\eneq

\beq
r^2e^{\frac{\lambda-\nu}{2}}f''(R)\frac{\partial\gamma}{\partial \lambda}=0,          \label{2.10}
\eneq

\beq
r^2e^{\frac{\lambda-\nu}{2}}f''(R)\frac{\partial\beta}{\partial R}=0,                 \label{2.11}
\eneq

\beq
-r^2e^{\frac{\nu-\lambda}{2}}f''(R)\frac{\partial\gamma}{\partial \lambda}=0,              \label{2.12}
\eneq

\beq
\frac{\alpha}{2}+\frac{\partial\alpha}{\partial\nu}-\frac{1}{2}\beta+\frac{f'''(R)}{f''(R)}\gamma+\frac{\partial\gamma}{\partial R}=0,           \label{2.13}
\eneq

\beq
-r^2e^{\frac{\nu-\lambda}{2}}f''(R)\frac{\partial\alpha}{\partial \lambda}=0,                    \label{2.14}
\eneq

\beq 
r^2e^{\frac{\lambda-\nu}{2}}f''(R)\frac{\partial\gamma}{\partial \nu}=0,          \label{2.15}                                  
\eneq

\beq
r^2e^{\frac{\lambda-\nu}{2}}f''(R)\frac{\partial\beta}{\partial \nu}=0,            \label{2.16}
\eneq

\beq
\frac{1}{2}e^{\nu-\lambda}\left(\beta-\alpha\right)+\frac{\partial\beta}{\partial\lambda}+\gamma\frac{f'''(R)}{f''(R)}+\frac{\partial\gamma}{\partial R}=0,      \label{2.17}
\eneq
where the following constraint has to be satisfied 

\bea \label{2.18}
&&\frac{1}{2}r^2\left(f(R)-Rf'(R)\right)\left(\alpha+\beta\right)-f'(R)\left(\alpha+\beta\right)-r^2Rf''(R)\gamma-2f''(R)\gamma+e^{-\lambda}\left[f'(R)\left(\alpha-\beta\right)+ \nonumber \right. \\
&&-r\lambda_rf'(R)\left(\alpha-\beta\right)-2r\lambda_rf'(R)\frac{\partial \beta}{\partial \lambda}+\left.2\gamma f''(R)\left(1-r\lambda_r\right)\right]=0.
\enea
If we consider that $f''(R)\neq0$, that is the trivial solution $f(R)=R$ is excluded,  Eqs. (\ref{2.8})-(\ref{2.12}) and (\ref{2.14})-(\ref{2.16}) can be immediately solved giving the result 

\beq
\alpha=\alpha(\nu)\,,\quad 
\beta=\beta(\lambda)\,,\quad
\gamma=\gamma(R)\,.
\eneq
Eq. (\ref{2.13}) can be recast in the form 

\beq
\frac{\alpha}{2}+\frac{d\alpha}{d\nu}-\frac{1}{2}\beta=-\frac{f'''(R)}{f''(R)}\gamma-\frac{d\gamma}{dR}=\mu_0, \label{2.19}
\eneq
where $\mu_0$ is a constant, because the r.h.s. of the first equality depends only on $R$ while the l.h.s. one is a function of $\nu$ and $\lambda$, which are independent variables. 
We can assume $\mu_0=0$, so  we obtain, from Eq. (\ref{2.19}) the following set of equations:

\beq
\frac{\alpha}{2}+\frac{d\alpha}{d\nu}=\frac{1}{2}\beta,     \label{2.20}
\eneq
 
\beq
\frac{d\gamma}{\gamma}=-\frac{f'''(R)}{f''(R)}dR.      \label{2.21}
\eneq
By the same arguments used above, we obtain, from Eq. (\ref{2.20}), that $\beta$ is constant, and using this result and putting Eq. (\ref{2.21}) into Eq. (\ref{2.17}),
we get

\beq
\alpha=\beta=\alpha_0,
\eneq 
where $\alpha_0$ is a constant, being this result consistent with Eq. (\ref{2.20}). 
Moreover, integrating Eq. (\ref{2.21}) we obtain

\beq
\gamma=\frac{\mu_1}{f''(R)},
\eneq
 being $\mu_1$ an integration constant. 
Now the constraint (\ref{2.18}) becomes

\beq \label{2.22}
\alpha_0\left(f(R)-Rf'(R)\right)r^2-2\alpha_0f'(R)-\mu_1r^2R-2\mu_1+2\mu_1e^{-\lambda}\left(1-r\lambda_r\right)=0.  
\eneq
Using Eq. (\ref{2.6}) for the conservation of the  current $j^\mu$, we immediately obtain 

\beq \label{2.23}
\partial_r\left[\left(r^2e^{\frac{\nu-\lambda}{2}}\right)\left(\alpha_0f''(R)R_r+\alpha_0\frac{2}{r}f'(R)+\mu_1\nu_r \right)\right]= 
\partial_t\left[\left(r^2e^{\frac{\lambda-\nu}{2}}\right)\left(\alpha_0f''(R)R_t+\mu_1\lambda_t \right)\right].    
\eneq
For  the particular choice $\alpha_0=0$, the constraint (\ref{2.22})  becomes

\beq
R=-\frac{2}{r^2}+e^{-\lambda}\left(\frac{2}{r^2}-\frac{2}{r}\lambda_r\right),             \label{2.24}
\eneq
while Eq. (\ref{2.23}) is

\beq
\partial_r\left(r^2e^{\frac{\nu-\lambda}{2}}\nu_r \right)=\partial_t\left(r^2e^{\frac{\lambda-\nu}{2}}\lambda_t \right).         \label{2.25}
\eneq
Using Eq. (\ref{2.25}) into the expression for the Ricci scalar (\ref{1.2}), it is immediate to verify that the dynamical constraint (\ref{2.24}) is satisfied. In this way we have that our system of equations is  consistent having the four equations (\ref{1.6}), (\ref{1.7}), (\ref{2.24}) and (\ref {2.25}), in the four unknown functions $\nu(r,t)$, $\lambda(r,t)$, $R(r,t)$ and $f(R)$.  Our task is now to determine these unknown functions selected by the presence of the Noether symmetry.

\section{Exact Solutions}              
\label{V}

We are able to solve the dynamics by the Noether symmetry approach. Considering Eqs. (\ref{1.6}), (\ref{1.7}) and (\ref{2.24}), in the case of constant curvature $R=R_0$, we obtain the following solutions for the metric coefficients

\beq
e^{\lambda(r,t)}=\frac{1}{1+\frac{R_0r^2}{6}-\frac{B_1(t)}{2r}}= B_2(t)e^{-\nu(r,t)},       \label{2.26}
\eneq
where $B_1(t)$ and $B_2(t)$ are integration functions. Moreover we get as constraint  $f(R_0)=0$. 
Then Eq. (\ref{2.25}) for the conserved current, using the solution (\ref{2.26})  with $B_1(t)=B_1$=constant, gives 

\beq
R_0=0.                                                                                        
 \label{2.27}
\eneq
Finally, using this result in Eq. (\ref{2.26}) and rescaling the time as in Eq. (\ref{1.14}), we obtain the following time-independent solution

\bea                                                                                           
ds^2=\left(1-\frac{B_1}{2r}\right)d\tilde{t}^2-\frac{1}{\left(1-\frac{B_1}{2r}\right)}dr^2-r^2d\Omega^2,   \label{2.28}
\enea 
which is the well-known  Schwarzschild solution in  GR.  We can  conclude that Noether Symmetry Approach for $f(R)$-gravity reproduces the same solutions of GR in the case of constant curvature. This is not a new result but it is useful to test the method  \cite{spherical}.

However, other interesting solutions can be found in the case of non-constant curvature in a spherically symmetric static metric. In fact,  the current conservation Eq. (\ref{2.25}) becomes now

\beq
\frac{d}{dr}\left(r^2e^{\frac{\nu-\lambda}{2}}\nu_r\right)=0,    \label{2.30}
\eneq
where $\nu=\nu(r)$ and $\lambda=\lambda(r)$. Assuming, as standard,  that $\nu=-\lambda$ \cite{chandra}, Eq. (\ref{2.30})  can be easily solved to give

\beq
\nu(r)=ln\left(C_1-\frac{1}{r}\right)+C_2=-\lambda(r),
\eneq
being $C_1$ and $C_2$ integration constants. Subtracting Eqs. (\ref{1.6}) and (\ref{1.7}), after some  algebra, we obtain

\beq
f'(R(r))=D_1r+D_2,        \label{2.29}
\eneq
where, again,  $D_1$ and $D_2$ are integration constants. In order to find the form of the $f(R)$-model, we need to express $f'(R)$ as an explicit function of $R$.
Then using the expression for the Ricci scalar (\ref{2.24}), which now is

\beq
R(r)=2\frac{\left(C_1e^{-C_2}-1\right)}{r^2},
\eneq
we obtain $r$ as a function of $R$. Substituting into Eq. (\ref{2.29}), we are able to perform the integration obtaining

\beq
f(R)=2D_1\left(2C_1e^{-C_2}-2\right)^{\frac{1}{2}}R^{\frac{1}{2}}+D_2R+D_3, 
\eneq
where $D_3$ is a constant.
It can be demonstrated  that our system of equations, considering also Eq. (\ref{1.7}), is solved for  given choices of the constants. For  $C_1=\frac{1}{2}$, $C_2=0$, $D_2=-3D_1$ and $D_3=0$, 
we get the metric
\beq
ds^2=\left(\frac{1}{2}-\frac{1}{r}\right)dt^2-\frac{1}{\left(\frac{1}{2}-\frac{1}{r}\right)}dr^2-r^2d\Omega^2. 
\eneq
The selected $f(R)$-function  is

\beq
f(R)=D_2R-\frac{2}{3}D_2\sqrt{-R}.
\eneq
The Ricci scalar is uniquely determined as

\beq
R=-\frac{1}{r^2}.
\eneq
This solution is physically consistent for $r>0$ and it is   asymptotically flat.

\section{Conclusions}       \label{VI}
In the context of $f(R)$-gravity,  a  Lagrangian approach has been developed  to study  dynamics of  spherically symmetric metrics. We have obtained the Euler-Lagrange equations and  solved them in the  case of constant curvature $R=R_0$ recovering the standard  Schwarzschild-de Sitter solution of GR. In this case, the Birkhoff theorem holds. Then the Noether Symmetry Approach has been developed in order to reduce  dynamics by  finding out conserved quantities, which can be  expressed as  currents.  After, we have solved  the Noether system related to the condition  $L_{\bf X}\mathcal{L}=0$, deriving a constraint equation over the dynamics and a conserved current expressed as  functions of the dynamical variables. The constraint allows to  select the form of the $f(R)$-model.  From this point of view the Noether symmetry ensures the closure of the dynamical system. Finally, we have presented some  particular cases showing  that
 Noether symmetries are compatible with GR and that solutions exist also for other $f(R)$-models different from $f(R)=R$.  In conclusion, the method reveals a useful approach both to select consistent $f(R)$-models and to find out exact solutions.

\end{document}